\documentclass[preprint,showpacs,preprintnumbers,amsmath,amssymb]{revtex4}

\usepackage{graphicx}
\usepackage{dcolumn}
\usepackage{bm}

\begin{document}

\title{Brane world regularization of point particle classical self-energy}

\author{Rom\'an Linares$^1$}
\email{lirr@xanum.uam.mx}
\author{Hugo A. Morales-T\'ecotl$^1$}
\email{hugo@xanum.uam.mx}
\author{Omar Pedraza$^2$}
\email{omarp@uaeh.edu.mx}
\author{Luis O. Pimentel$^1$}
\email{lopr@xanum.uam.mx}

\affiliation{$^1$ Departamento de F\'{\i}sica, Universidad Aut\'onoma Metropolitana Iztapalapa,\\
San Rafael Atlixco 186, C.P. 09340, M\'exico D.F., M\'exico,}

\affiliation{$^2$
\'Area Acad\'emica de Matem\'aticas y F\'isica, Universidad Aut\'onoma del Estado de Hidalgo, \\
Carretera Pachuca-Tulancingo Km. 4.5, C P. 42184, Pachuca,
M\'exico.}


\begin{abstract}
Physical effects in brane worlds models emerge by the incorporation of
field modes coming from extra dimensions with the usual
four dimensional ones. Such effects can be tested with well established experiments to set bounds on the parameters of the brane models.
In this work we extend a previous result which gave finite electromagnetic potentials and self energies for a source looking pointlike to an observer sitting in a 4D Minkowski subspace of a single brane of a Randall-Sundrum spacetime including compact dimensions, and along which the source stretches uniformly. We show that a scalar particle produces a nonsingular static potential, possess a finite self-energy and that technically its analysis is very similar to  the electrostatic case. As for the latter, we use the deviations from the Coulomb potential to set bounds on the anti de Sitter radius of the brane model on the basis of two experiments, namely, one of the Cavendish type and other being the scattering of electrons by Helium atoms. We found these are less stringent than others previously obtained using the Lamb shift in Hydrogen.
\end{abstract}

\pacs{11.25.Wx, 11.10Kk, 11.25.Mj,04.25.-g, 03.50.-z}
\maketitle
\section{Introduction}

Despite the extraordinarily rich accuracy with which the predictions
of electrodynamics have been experimentally tested over the years
(see for instance \cite{Goldhaber:2008xy,Goldhaber:1971mr} and
references therein), efforts to place limits on deviations from its
standard formulation continue nowadays. The nature of the
experiments cover a big range of possibilities which include among
others: a) Testing the power in the inverse-square law of Coulomb, b)
Seeking a nonzero value for the rest mass of the photon and c)
Considering more degrees of freedom, allowing mass for the photon
while preserving explicit gauge invariance. It is worth to mention
that all these experiments have probed length scales increasing
dramatically over time.

Now, historically, once the Maxwell theory of electromagnetic fields
was established, one of the main concerns in physics was the
construction of a consistent description of electrodynamics and
charged particles. The first serious proposals in this direction
were developed by Lorentz \cite{Lorentz} and Abraham \cite{Abraham}.
These proposals and subsequent attempts based on classical
electrodynamics, special relativity and the Lorentz force law led to
the theory known as Classical Electron Theory (CET). Whereas in the
Maxwell theory charges are considered as punctual which produce
infinite Lorentz self-force and infinite electromagnetic
self-energies associated with the singularities of the
Li\'enard-Wiechert potential, in the CET charges are considered as
extended objects that experience a volume-averaged Lorentz force.
Parallel to the development of the CET, quantum mechanics was
developed giving origin to one of the most spectacular theory we
have in physics: Quantum Electrodynamics (QED). This theory is
awesome due to the impressive range of electromagnetic phenomena it
covers with spectacular precision. After the experimental
achievements of QED, CET dropped from the list of contenders for a
fundamental theory of electrodynamics interacting with matter.
However despite the great success of QED there are still some
features of the theory that could be waiting for a better
explanation, for instance, its property of renormalizability. It
turns out that QED is defined by a perturbative series that is
renormalizable in each order, but it is most likely to be merely
asymptotic in character rather than convergent \cite{Dyson:1952tj},
in such a way that the precision results are obtained only when
computations are made to some order in the expansion series, but
without any a priori prescription to stop the series at some order.

It is thus tempting to investigate theories that avoid singularities.
These are not expected to solve the problem but at least they may contribute to a better understanding of the singularity issue.

In this context, recently, in
a previous work, some of us found that a source lying on the single
brane of a Randall-Sundrum spacetime including compact dimensions,
and which looks pointlike to an observer sitting in usual 3D space,
produces a static potential which is non singular at 3D point
position. Moreover it matches Coulomb's potential outside a small
neighborhood \cite{MoralesTecotl:2006eh}. The presence of the
compact dimensions in this setup serve to localize the gauge field
on the brane \cite{Dubovsky:2000am,Dubovsky:2000av,Oda:2000zc}. The
aim of this paper is to investigate further some consequences of the
above property to set bounds to the AdS curvature radius $\epsilon$
using the experimental results from the Cavendish experiment for
electromagnetism and the scattering of electrons by Helium atoms.
For the sake of clarity, the simpler case of a scalar particle is
first considered. Remarkably, the nonsingular character of the
potential holds together with the finiteness of the selfenergy.
Indeed, technically, the study of the potentials for both scalar and
electromagnetic is very similar.

Our interest in this work is twofold, on one side it is interesting
to explore how the old problem of divergences acquires a different
character in light of the brane world models, at least classically,
and, on the other hand, it is also interesting from the perspective
of the brane-world scenarios
\cite{Randall:1999vf,Randall:1999ee,Antoniadis:1998ig,Arkani-Hamed:1998rs,Antoniadis:1990ew},
which have recently been matter of a copious research, mainly in
high energy physics (see e.g. \cite{Allanach:2004ub,Csaki:2004ay},
and references therein) and cosmology (see e.g.
\cite{Maartens:2003tw,Elizalde:2006iu,Maartens:2010ar}, and
references therein). More recently, the possibility to obtain
information from models with extra dimensions studying low energy
physical phenomena has also been addressed. In particular we mention
the ones that have been performed in the RSII-$p$ setup, such as the
electric charge conservation \cite{Dubovsky:2000av}, the Casimir
effect between parallel plates \cite{Linares:2010uy,Frank:2008dt}
and the Hydrogen Lamb shift \cite{MoralesTecotl:2006eh}.

The paper is organized as follows. In section \ref{background} we
briefly describe the RSII-$p$ setup, section \ref{ScalarCase} is
dedicated to obtain the static potential for a scalar field whereas
in section \ref{ElectCase} we do the same for the electric case. In
section \ref{SubsecCavendish} we set bounds to the AdS radius
$\epsilon$ comparing our electrostatic results with the experimental
values obtained in Cavendish like experiments of the Coulomb force.
Section \ref{SubsecHelium} is dedicated to the same purpose but this
time we use the experimental results of the scattering process of
electrons by  Helium atoms. Finally section
\ref{conclusions} is devoted to a brief discussion.

\section{Randall-Sundrum II-$p$ scenarios}\label{background}

The Randall-Sundrum II-$p$ scenarios consist of a ($3+p$)-brane with
$p$ compact dimensions and positive tension $\sigma$, embedded in a
($5+p$) spacetime whose metrics are two patches of anti-de Sitter
(AdS$_{5+p}$) having curvature radius $\epsilon$. The interest in
these models comes from its property of localizing on the brane:
scalar, gauge and gravity fields due to the gravity produced by the
brane itself. This property is valid whenever there are $p$ extra
compact dimensions \cite{Dubovsky:2000av,Oda:2000zc}. In the
limiting case $p=0$, the model only localizes scalar and gravity
fields. With this setup and appropriate fine-tuning between the
brane tension $\sigma$ and the bulk cosmological constant $\Lambda$,
which are related to $\epsilon$ as follows
\begin{equation}
\sigma=\frac{2(3+p)}{8\pi \epsilon G_{5+p}}, \quad
\Lambda=-\frac{(3+p)(4+p)}{16\pi \epsilon^2
G_{5+p}}=-\frac{(4+p)\sigma}{4\epsilon},
\end{equation}
there exists a solution to (5+$p$)D Einstein equations with metric
\begin{equation}\label{mebra}
ds_{5+p}^{\, 2} = e^{-2 |y|/\epsilon}
\left[\eta_{\mu\nu}dx^{\mu}dx^{\nu}-\sum_{i=1}^pR^2_id\theta_i^2
\right]-dy^2.
\end{equation}
Here $\eta_{\mu\nu}$ is the 4D Minkowski tensor, $\theta_i \in
[0,2\pi]$ are $p$ compact coordinates, $R_i$ are the sizes of
compact dimensions, $G_{5+p}$ is the ($5+p$)D Newton constant.
Throughout the paper we will use the following notation for the
5+$p$ coordinates $X^M\equiv (x^\mu, \theta_i , y)$, where
$\mu=0,1,2,3$, and $i=1, \dots p$.

In this work we consider two different $(5+p)D$ field theories on
RSII-$p$: a massless scalar field and electrodynamics. They will be
subjected to a hybrid of the two well known consistent compactifications,
namely Kaluza-Klein (KK) \cite{Kaluza:1921tu,Klein:1926tv} and
warped \cite{Randall:1999vf}. These two differ among them on whether
the compactified manifold is factorizable or not. The corresponding
effective field theories in 4D Minkowski space-time will be given.

In regard to the KK compactification, it is well known toroidal dimensional
reductions lead to consistent lower dimensional theories which nonetheless
can be questioned in that they do not come with a mechanism to fix the
moduli, or equivalently, the radii of the $p$D torus $T^p$ \cite{Duff:1986hr,Green:1987mn}.
Historically, a way out in such cases, has been to conform with the
corresponding phenomenology at low  enough energies and set a bound
for the radii (e.g. the use of the classical value of the electron charge
required a radius of the order Planck length in the original KK
setting  \cite{Klein:1926tv,Klein:1926fj}). We will adhere to this
approach by considering a low energy approximation so that we truncate the
massive KK modes of the compact dimensions but keep those corresponding to the
noncompact dimension just meaning that we assume the energy scale of the
former is much smaller than that of the latter.  This is explicitly performed
in the Green's function in \ref{ScalarCaseA} for the scalar field
and in \ref{ElectCase} for the gauge field. As for the consistency of the
Randall-Sundrum compactification it has been discussed in  \cite{Rubakov:2001kp} (and references therein).

For completeness we only mention other mechanisms adopted in the literature to perform a generalized KK compactification.
One of them is the so called Scherk-Schwarz compactification
\cite{Scherk:1978ta,Scherk:1979zr} or flux compactifications
\cite{Grana:2005jc,Blumenhagen:2006ci,Douglas:2006es}. In this
mechanism the symmetries of the compactification manifold and/or the
fields are used to produce an effective potential for stabilizing
the size of the extra dimensions. There also exists a quantum
proposal by Candelas and Weinberg \cite{Candelas:1983ae} where the
effective potential for the moduli fields is produced  by the
Casimir energy of matter fields or gravity. It remains open to
study these possibilities for our present setup.

A remark regarding the stability of the
scenario described by the metric (\ref{mebra}) is here in order.
Concerning the world volume of the (3+$p$)-brane,
$M_4 \otimes T^p$, it is clear that the space is stable since it is flat.
On the other hand the stability of the space-time
(\ref{mebra}), without the $T^p$ structure, was studied
long ago in \cite{Randall:1999vf,Garriga:1999yh} for static
perturbations of the metric and in \cite{Sasaki:1999mi} for general
space-time dependent sources. The stability of other warped compactifications has
also been addressed, for instance in
\cite{Goldberger:1999uk,Lesgourgues:2003mi,Maity:2006ub,Das:2007mr}
it was discussed the moduli stabilization of the RSI model whereas
in \cite{Flachi:2003bb} it was discussed for more general metrics.

Before ending this section is worth to mention that this setup has
been considered in different low energy physics effects such as the
electric charge conservation \cite{Dubovsky:2000av}, the Casimir
effect between two conductor hyperplates
\cite{Linares:2007yz,Linares:2008am,Frank:2008dt,Linares:2010uy} and
the Liennard-Wiechert potentials and Hydrogen Lamb shift
\cite{MoralesTecotl:2006eh} among others.

\section{Static potential for a scalar field}\label{ScalarCase}

\begin{figure}[htb]
\center{\includegraphics[height=5cm,width=7cm]{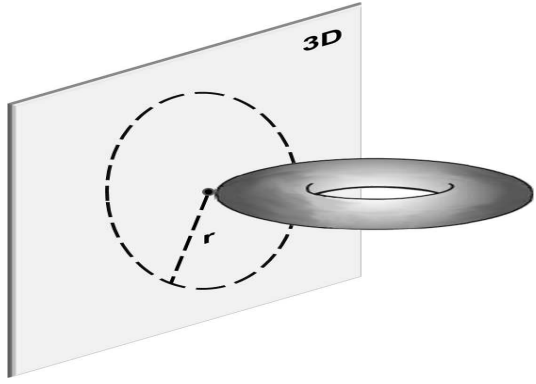}}
\caption{Schematic view of the charge source for $p$=2. The source
is effectively pointlike from the perspective of an observer sitting
in the usual 3d space.} \label{brana}
\end{figure}

In this section we compute the potential produced by a static source
which is seen as punctual by an observer living on the usual 3D
subspace of the (3+$p$)-brane. It stretches however along the $p$
compact dimensions thus forming a $p$-dimensional torus
(\ref{brana}).

\subsection{The Green's function}\label{ScalarCaseA}
Let us consider a massless scalar field $\Phi$ described by the
action in (5+$p$)D
\begin{equation}\label{ActionScalar}
S= \int \, d{}\,^4x  \,  \prod_{i=1}^p \, R_i d \theta_i \, dy\,
\sqrt{|g|} \, \left( \frac12 g^{MN}\partial_M \Phi \,
\partial_N \Phi + \Phi J_{scalar} \right).
\end{equation}
The equation of motion for the scalar field is
\begin{equation}\label{ec.escalar}
\frac{1}{\sqrt{|g|}}
\partial_M \left( \sqrt{|g|}g^{MN}\partial_N\Phi \right) = J_{scalar}
\end{equation}
where the source is given by
\begin{equation}\label{Jscalar}
J_{scalar} = \lambda^{(p)} \delta^3 \left(\vec{x}-\vec{x}_0\right)
\delta \left(y-y_0\right).
\end{equation}
Here $\lambda^{(p)}$ is a constant whose dimensions are
[charge]/[length]$^{p}$, explicitly: $\lambda^{(p)}=
\frac{\lambda}{(2\pi)^p R_1 \cdots R_p}$, with $\lambda$ the total
charge.

In the background (\ref{mebra}), the equation of motion
(\ref{ec.escalar}) becomes
\begin{equation}
e^{2|y|/\epsilon} \left[ \Box \Phi - \sum_{i=1}^p \frac{1}{R_i^2}\,
\partial_{\theta_i}^{\, 2} \Phi \right] - \frac{1}{\sqrt{|g|}}\,
\partial_{y} \left[\sqrt{|g|}
\partial_{y} \Phi \right] = J_{scalar},
\end{equation}
where $\Box$ stands for the flat 4D D'Alambertian. The corresponding
Green's equation is
\begin{equation}
e^{2|y|/\epsilon} \left[ \Box G - \sum_{i=1}^p \frac{1}{R_i^2}\,
\partial_{\theta_i}^{\, 2}G \right] - \frac{1}{\sqrt{|g|}}\,
\partial_{y}\left[\sqrt{|g|}
\partial_{y}G\right] = \frac{\delta(y-y') \delta^p(R_i\theta_{i}-R_i\theta'_{i})
\delta^4(x-x')}{\sqrt{|g|}},
\end{equation}
where $G$ is the $(5+p)$D Green's function. This can be expressed in
terms of the eigenfunctions of the differential operators for the
different coordinates. Assuming  $\Psi(X^M) \equiv e^{ik_\mu x^\mu}
\prod_{i=1}^p \Theta_i(\theta_i) \psi(y)$, where the modes
$\Theta_n$ and $\psi_m$ account for the $\theta$ and $y$ dependence
respectively. These have been discussed previously (see for instance
\cite{Dubovsky:2000am,Linares:2010uy}) and here we only give a
summary. The differential equations governing the $p$ compact modes
depending on $\theta_i$ are
\begin{equation}\label{ThetaEq}
(\partial_{\theta_i}^2+m_{\theta_i}^2 R_i^2)\Theta_i(\theta_i)=0,
\hspace{1cm} i=1,\cdots,p,
\end{equation}
whereas for the noncompact modes depending on $y$ one gets
\begin{equation}\label{Yequation}
(\partial_y^2-\frac{(4+p)}{\epsilon} \mbox{sgn}(y)
\partial_y)+m^2e^{2 |y|/\epsilon}) \psi(y)=0.
\end{equation}
The $(p+1)$ constants of separation, $m_{\theta_i},m$, fulfill the
following dispersion relation
\begin{equation}
k^2= \sum_{i=1}^p m_{\theta_i}^2+m^2 \equiv m_p^2+m^2.
\end{equation}
To account for the compactness of the $p$ dimensions Eq.
(\ref{ThetaEq}) is solved under the periodic boundary conditions
\begin{equation} \label{thetaperiodic}
\Theta_{n_i}(\theta_i)=\Theta_{n_i}(\theta_i+2\pi),
\end{equation}
and the solutions turn out to be
\begin{equation}\label{modesT}
\Theta_{n_i}(\theta_i)= \frac{1}{\sqrt{2\pi R_i}} e^{in_i \theta_i},
\, \, \, \mbox{where} \, \, \, n_i = m_{\theta_i}R_i \in \mathbb{Z}.
\end{equation}
To match the modes across the brane along the non-compact dimension,
equation (\ref{Yequation}) is solved with the following boundary
conditions
\begin{equation}\label{BCbrane}
\psi(y=0^+)=\psi(y=0^-) \qquad {\rm and} \qquad
\partial_y \psi(y=0^+)= \partial_y \psi(y=0^-).
\end{equation}
In this case the solutions include a massless zero mode localized on
the brane
\begin{equation}\label{MasslessY}
\psi_0(y)=\sqrt{\frac{2+p}{2\epsilon}}
\end{equation}
which satisfies the normalization condition $2 \int_0^{\infty} dy
e^{-(2+p)|y|/\epsilon} \psi_0^2(y)=1$, as well as massive modes
given by
\begin{equation}\label{MassiveY}
\psi_{m}(y)=e^{\gamma y / \epsilon} \sqrt{\frac{m
\epsilon}{2}}\left[a_m J_{\gamma}\left(m \epsilon \, e^{y/
\epsilon}\right)+b_mN_{\gamma}\left(m \epsilon \,
e^{y/\epsilon}\right) \right],
\end{equation}
where $J_{\gamma}$ and $N_{\gamma}$ are the Bessel and Neumann
functions, respectively. In this expression
\begin{equation}\label{Gammap}
\gamma \equiv \frac{4+p}{2},
\end{equation}
and the coefficients $a_m$ y $b_m$ are
\begin{equation}\label{constants}
a_m = -\frac{A_m}{\sqrt{1+A_m^2}}, \quad b_m =
\frac{1}{\sqrt{1+A_m^2}}, \quad A_m = \frac{N_{\gamma-1} \left(m
\epsilon\right)}{J_{\gamma-1}\left(m \epsilon \right)}.
\end{equation}
Notice that in this case, the localization of the massive modes on
the brane is better for increasing $p$ , since the modes are
modulated exponentially by a factor of $e^{-p|y|/ (2 \epsilon) }$.
The normalization condition for the massive modes is
$\int_{-\infty}^\infty dy e^{-(p+2)|y|/ \epsilon }\psi_m(y)
\psi_{m'}(y)=\delta(m-m')$.

With the eigenfunctions at hand it is straightforward to use them to
write down the Green's function. It takes the form
\begin{equation}\label{FormalGreen}
G(x,\theta_i,y;x',\theta'_i,y')=\prod_{i=1}^p\sum_{\{n\}}\frac{e^{in_i\theta_i}e^{-in_i\theta_i'}}{2\pi
R_i}\int \frac{d^4k}{(2\pi)^4} e^{ik_{\mu}(x^{\mu}-x'^{\mu})} \left[
\frac{\psi_0(y)\psi_0(y')}{k^2-m_p^2}+\int_0^{\infty}dm
\frac{\psi_m(y)\psi_m(y')}{k^2-m^2-m_p^2} \right],
\end{equation}
where $\{n\}$ denotes $\{n_1,n_2,\dots,n_p \, | \, n_1\in
\mathbb{Z},\dots,n_p\in \mathbb{Z}\}$.

At this point it is convenient to introduce an approximation that
will allow us to obtain analytic expressions of the potential. We
have massive modes from both compact dimensions (\ref{modesT}) and
the noncompact one (\ref{MassiveY}). Since we are interested in the
low energy regime we assume $m_{\theta_i} \ll m \ll \epsilon^{-1}$.
Hence we will set $n_1=\dots=n_p=0$, and, as for the noncompact
modes (\ref{MassiveY}) we use
\begin{equation}\label{MassiveLow}
\psi_m(y) \approx -e^{\gamma y/ \epsilon} \sqrt{\frac{m
\epsilon}{2}} J_{\gamma} \left(m \epsilon \, e^{y/ \epsilon
}\right).
\end{equation}
In such low energy regime and upon integrating over the $p$ compact
extra dimensions, we end up with an effective 5D Green's function
\begin{equation}\label{LowEnergyGreen}
G_{5D}(x,y;x',y') = \int \frac{d^4k}{(2\pi)^4}
e^{ik_{\mu}(x^{\mu}-x'^{\mu})} \left[
\frac{\psi_0(y)\psi_0(y')}{k^2}+\int_0^{\infty}dm
\frac{\psi_m(y)\psi_m(y')}{k^2-m^2} \right],
\end{equation}
where the massless mode is given by (\ref{MasslessY}) and the
massive modes by (\ref{MassiveLow}). Although we have only taken the
zero modes of the compact extra dimensions, notice that their
imprints remain in the 5D Green's function through (\ref{Gammap})
and (\ref{MassiveLow}).

\subsection{Static potential}

Now we are in position to compute the static potential. In this case
the useful Green's function is
\begin{eqnarray}
G(\vec{x},y;\vec{x'},y')&=&\int_{-\infty}^{\infty}dt'G(\vec{x},t=0,y;\vec{x'},t',y')\nonumber\\
&=&\frac{\psi_0(y)\psi_0(y')}{4\pi r}+\int_0^{\infty}dm
\psi_m(y)\psi_m(y') \frac{e^{-mr}}{4\pi r}
\end{eqnarray}
where $r=|\vec{x}-\vec{x}'|$. As usual the potential is obtained
upon integrating the Green's function times the source, Eq.
(\ref{Jscalar}), and we are interested in its form at the brane,
i.e. $y=0$, namely,
\begin{eqnarray}
\varphi(r,y=0)&=&\int d^3x'dy' G(\vec{x},y=0;\vec{x'},y')\, J_{scalar}(\vec{x'},y';\vec{x_0},y_0) \nonumber \\
&=& \frac{\lambda^{(p)}\psi_0(0)\psi_0(y_0)}{4\pi
r}+\lambda^{(p)}\int_0^{\infty}dm \psi_m(0)\psi_m(y_0)
\frac{e^{-mr}}{4\pi r}, \label{poe}
\end{eqnarray}
where now $r=|\vec{x}-\vec{x}_0|$,  and (\ref{MassiveLow}) takes the
asymptotic value
\begin{equation}
\psi_m(0) \approx \frac{1}{\Gamma(\gamma - 1)}
\sqrt{\frac{m\epsilon}{2}}
\left(\frac{m\epsilon}{2}\right)^{\gamma-2} . \label{ecm1}
\end{equation}
Finally, the potential becomes
\begin{equation}
\varphi(r) = \frac{\lambda^{(p)}}{4\pi
r}\left(\frac{2+p}{2\epsilon}\right) - \lambda^{(p)}
\int_0^{\infty}dm \frac{1}{\Gamma(\gamma-1)}
\left(\frac{m\epsilon}{2}\right)^{\gamma-1} e^{\gamma y_0/\epsilon}
J_{\gamma}\left({m}{\epsilon}e^{y_0/\epsilon}\right)\frac{e^{-mr}}{4\pi
r}. \label{ecpoe}
\end{equation}
Next we further assume the source to be located at the brane, i.e.
$y_0 = 0$. The explicit form of (\ref{ecpoe}) now depends on whether
the number of extra compact dimensions, $p$, is odd or even, and so
we discuss each case separately.

\subsection{Odd number of extra compact dimensions}

In the case that $p$ takes odd values, $\gamma$ takes semi-integer
values and is useful to use the relation
\begin{equation}
m^{l+1/2}J_{l+1/2}\left(m\epsilon\right)=(-1)^l\sqrt{\frac{2}{\pi}}\epsilon^{l+1/2}
\left(\frac{d}{\epsilon
d\epsilon}\right)^l\frac{\sin(m\epsilon)}{\epsilon}\label{re1},
\end{equation}
in the integrand of (\ref{ecpoe}). Upon evaluation of the integral
we get
\begin{equation}
\varphi(r)=\frac{\lambda^{(p)}}{4\pi r}\frac{2+p}{2\epsilon}-
\frac{(-1)^{\gamma-\frac{1}{2}}\lambda^{(p)}\epsilon^{2\gamma-1}
}{2^{\gamma-3/2}\sqrt{\pi}\Gamma(\gamma-1)} \frac{1}{4\pi
r}\left(\frac{d}{\epsilon d\epsilon}\right)^{\gamma-\frac{1}{2}}
\left(\frac{\pi}{2\epsilon}-\frac{\arctan\left(\frac{r}{\epsilon}\right)}{\epsilon}\right).
\end{equation}
Let us notice the first term of this expression is divergent at $r=0$. However such a term cancels out
with the first term within parenthesis for every odd $p$. This
leads to the form of the effective potential
\begin{equation}\label{potscalarodd}
\varphi(r)= \frac{(-1)^{\gamma-\frac{1}{2}}
\lambda^{(p)}\epsilon^{2\gamma-1}
}{2^{\gamma-3/2}\sqrt{\pi}\Gamma(\gamma-1)} \frac{1}{4\pi
r}\left(\frac{d}{\epsilon d\epsilon}\right)^{\gamma-\frac{1}{2}}
\left( \frac{\arctan\left(\frac{r}{\epsilon}\right)}{\epsilon}
\right).
\end{equation}
As an example, let us work out the case in which we have only one
compact extra dimension, {\em ie} $p=1 \, \Rightarrow \,
\gamma=5/2$. From (\ref{potscalarodd}) we obtain
\begin{equation}\label{varphip1}
\varphi(r) = \frac{2q_s^{(1)}}{3\pi\epsilon}
\left[3\frac{\arctan\left(\frac{r}{\epsilon}\right)}{\frac{r}{\epsilon}}
+ \frac{5}{\left(1+\frac{r^2}{\epsilon^2}\right)} +
2\frac{\frac{r^2}{\epsilon^2}}{\left(1+\frac{r^2}{\epsilon^2}\right)^2}\right],
\end{equation}
where $q_s^{(1)}=\frac{3\lambda^{(1)}}{8\pi\epsilon}$. The finite
value of the potential at the 3D point position of the source takes
the value
\begin{equation}
\lim_{r\rightarrow 0}\varphi(r)=\frac{16q_s^{(1)}}{3\pi\epsilon},
\end{equation}
evidently regularized by the existence of $\epsilon$ and $R$.
Using (\ref{varphip1}) we can compute the effective self-energy of the
point charge, as determined by a 3D observer
\begin{equation}
E_{self}^{(p=1)}:=\frac{1}{2}\int_{\mathbb{R}^3}d^3x \left(\nabla\varphi\right)^2 =\frac{85\left(q_s^{(1)}\right)^2}{9\epsilon}.
\end{equation}

\subsection{Even number of extra compact dimensions}

In the case that $p$ takes even values $\gamma$ is integer and we
can use the relation
\begin{equation}
m^{l}J_{l}(m\epsilon) = (-1)^l \epsilon^l \left( \frac{d}{\epsilon
d\epsilon} \right)^{l-1}
\left(-\frac{J_1(m\epsilon)m}{\epsilon}\right),\label{re2}
\end{equation}
in (\ref{ecpoe}), to obtain \footnote{It is worth
mentioning here an alternative approach to get the same results for
both odd an even $p$. It amounts to using the completeness relation
of the noncompact modes. In such a case the coefficient in front of
$1/r$ is proportional to the square of a Dirac delta and by using
either dimensional regularization or distribution operations the
coefficient vanishes thus obtaining the same result, {\em ie} that
the potential is finite at the 3D position of the source
\cite{MoralesTecotl:2006eh}.}
\begin{equation}\label{potscalareven}
\varphi(r) = \frac{(-1)^{\gamma+1} \lambda^{(p)}
\epsilon^{2\gamma-1}}{2^{\gamma-1} \Gamma(\gamma-1)}\frac{1}{4\pi
r}\left(\frac{d}{\epsilon d\epsilon}\right)^{\gamma-1}\left(
\frac{r}{\epsilon^2\sqrt{r^2+\epsilon^2}} \right).
\end{equation}
As an example, let us consider the lowest even value for $p$: $p=2
\, \Rightarrow \gamma=3$. From (\ref{potscalareven}) we obtain
\begin{equation}
\varphi(r) = \frac{q_s^{(2)}}{8\epsilon}
\left[\frac{8}{\sqrt{1+\frac{r^2}{\epsilon^2}}} +
\frac{4}{\left(1+\frac{r^2}{\epsilon^2}\right)^{\frac{3}{2}}} +
\frac{3}{\left(1+\frac{r^2}{\epsilon^2}\right)^{\frac{5}{2}}}\right],
\end{equation}
where  $q_s^{(2)}=\frac{\lambda^{(2)}}{2\pi\epsilon}$. In this case
the finite value of the potential at the 3D position of the source
is
\begin{equation}
\lim_{r\rightarrow 0}\varphi(r)=\frac{15q_s^{(2)}}{8\epsilon},
\end{equation}
whereas the source self-energy is given by
\begin{equation}
E_{self}^{(p=2)}=\frac{51975\pi^2\left(q_s^{(2)}\right)^2}{65536\epsilon}.
\end{equation}

\section{Electrostatic potential}\label{ElectCase}

The procedure to compute this potential is similar to the one we
used in the scalar case. In the spirit of avoiding repetition, we
describe briefly the computation giving special emphasis to the
aspects that are different with respect to the scalar case. A
previous discussion of the photon Green's function analysis in the
RSII-$p$ scenario can be found in \cite{Dubovsky:2000av}. We begin
by considering the (5+$p$)D action
\begin{equation}\label{ActionGauge}
S= \int \, d{}\,^4x  \,  \prod_{i=1}^p \, R_i d \theta_i \, dy\,
\sqrt{|g|} \, \left( \frac14 g^{MN} g^{PQ} F_{MP}F_{NQ} + A_M
J^N_{gauge} \right),
\end{equation}
leading to the equation of motion
\begin{eqnarray}\label{eq:Max}
\frac{1}{\sqrt{|g|}}
\partial_{M}\bigg(\sqrt{|g|}\,\,g^{MP}g^{NQ}F_{PQ}\bigg)=- j^{N}_{gauge}.
\end{eqnarray}
We consider a static source along the brane, uniformly distributed
along the $p$ extra compact dimensions, namely
\begin{eqnarray}
\sqrt{|g|}j_{gauge}^{N}=\rho^{(p)} \delta^{N}_0\delta^{3}(\vec{
x}-\vec{x}_0)\delta(y-y_0),\label{eq:jmu}
\end{eqnarray}
where $\rho^{(p)}$ is the charge density. Now we write down the
equation of motion for the gauge field in the background
(\ref{mebra}). In order to do this, it is convenient to fix the
gauge $A^{y}=0$ and $A^{\theta_i}=0$, which is consistent with the
value $J^{\theta_i}=0$ for the components of the current density in
the directions of the compact extra dimensions. Thus Eq.
(\ref{eq:Max}) becomes
\begin{equation}
{\cal O}\hat A^{\sigma}
-e^{-p|y|/\epsilon}\partial^{\sigma}\partial_{\mu}\hat A^{\mu} =
-R^{-p} e^{p|y|/\epsilon}\sqrt{g}\,\,j^{\sigma},\label{eq:MaxRSn}
\end{equation}
where we assume equal size compact dimensions, $R_i=R,i=1,\dots,p$,
and the differential operator $\cal O$ is defined as
\begin{equation}
{\cal O}:=e^{-(p+2)|y|/\epsilon}\left( -\partial^2_{y}+
\frac{p+2}{\epsilon}\, sgn(y)
\partial_{y}+e^{2|y|/\epsilon}\Box \right), \label{eq:OMax}
\end{equation}
and $\hat A^{\nu}=\eta^{\nu\mu}A_{\mu}$. Inspection of equation
(\ref{eq:MaxRSn}) reveals the term $\partial_{\mu}\hat A^{\mu}$ is
pure gauge on the brane, so we drop it from now on
\cite{Dubovsky:2000av}.

To solve (\ref{eq:MaxRSn}) let us notice that the differential
operator (\ref{eq:OMax}) is invariant under the change $y\to -y$, so
the solutions will inherit such symmetry. This is important since we
are looking for the potential on the brane. We shall adopt again the
Green's function method. As in the scalar case, the necessary tools
are the eigenfunctions and eigenvalues of the differential equation.

The eigenfunctions and eigenvalues for the $p$ compact modes are the
same as those for the scalar field, Eq. (\ref{modesT}). As for the
noncompact modes depending upon $y$ and subject to the boundary
conditions (\ref{BCbrane}) they fulfill again a Bessel equation and
have the following  form
\begin{equation}
\phi_0=\sqrt{\frac{p}{2\epsilon}},\quad \quad \phi_m (y)=e^{\nu
y/\epsilon}\sqrt{\frac{m\epsilon}{2}} \left[ a_m J_{\nu} \left(
{m\epsilon}\, e^{ y/\epsilon}\right) + b_m N_{\nu}\left(
{m\epsilon}\, e^{y/ \epsilon}\right) \right],
\end{equation}
where \begin{equation} \nu \equiv \frac{p+2}{2},
\end{equation}
and the constants $a_m$, $b_m$ are defined as in (\ref{constants}),
with $\gamma$ replaced by $\nu$. The modes are normalized in the
form $\int_{-\infty}^{\infty}dy\,e^{-p|y|/\epsilon}\phi_0^2 =1$ and
$ \int_{-\infty}^{\infty}dy
e^{-p|y|/\epsilon}\phi_m(m\epsilon)\phi_{m'}(m'\epsilon)=\delta(m-m')\,$.
Formally the Green's function, its low energy approximation and the
static potential on the brane are obtained from (\ref{FormalGreen}),
(\ref{LowEnergyGreen}) and (\ref{ecpoe}), replacing the scalar modes
$\psi$ by the gauge modes $\phi$ as well as  the factor $\gamma$ by
$\nu$. The electrostatic Green's function on the brane takes the
form
\begin{equation}\label{eq:green1}
G_{gauge}(\vec{x}-\vec{x}',y=0,y')= \frac{p}{2\epsilon}\frac{1}{4\pi
r}- \frac{1}{4\pi r}\frac{ e^{\nu
y'/\epsilon}}{\Gamma(\nu-1)}\left(\frac{\epsilon}{2}\right)^{\nu-1}\int_0^{\infty}dm\,
m^{\nu-1}J_{\nu}\left(my'\right) e^{-mr} \,.
\end{equation}
Since we are interested in the potential for a source located on the
brane, we have to evaluate the above expression in the limit $y'
\rightarrow 0$. As in the scalar case this limit is different
depending on whether $p$ is either even or odd. They are given explicitly below.

\subsubsection{$p$ odd}
In this case the potential gets the form
\begin{equation}\label{eq:A01}
A^0(r)= \frac{\sigma^{(5+p)}}{4\pi R^{p}r} \sqrt{\frac{2}{\pi}}
\frac{(-1)^{\nu}\epsilon^{2\nu-1}}{\Gamma(\nu-1)(2)^{\nu-1}}\left(\frac{d}{\epsilon
d\epsilon}\right)^{\nu-\frac12}
\left[\frac{\arctan\left(\frac{r}{\epsilon}\right)}{\epsilon}
\right] \,,
\end{equation}
where $r=|\vec{x}-\vec{x}_0|$. As an example notice that for one
extra compact dimension $p=1$ one gets
\begin{equation}\label{A0p1}
A^0(r)=\frac{2e}{\epsilon\pi}
\left(\frac{1}{1+\frac{r^2}{\epsilon^2}} +
\frac{\arctan\left(\frac{r}{\epsilon}\right)}{\frac{r}{\epsilon}}\right),\quad
e=\frac{e^{(6)}}{2R\epsilon^2}\,,
\end{equation}
which reduces to the Coulomb potential for $r\gg\epsilon$ and is
finite at the 3D source position
\begin{equation}
\lim_{r\rightarrow 0}A^0(r)=\frac{4e}{\pi\epsilon}.
\end{equation}
The self-energy in this case is
\begin{equation}
E_{self}^{(p=1)}:=\frac{1}{2}\int_{\mathbb{R}^3}d^3x \left(\nabla A^0\right)^2=\frac{5e^2}{32\pi^3\epsilon}.
\end{equation}
\subsubsection{$p$ even}

Now $\nu$ is an integer and
\begin{equation}\label{eq:A0ppar}
A^0(r)=\frac{(-1)^{\nu}\sigma^{(5+p)}\epsilon^{2\nu-1}}{2^{\nu-1}\Gamma(\nu-1)R^p}\frac{1}{4\pi
r}\left(\frac{d}{\epsilon d\epsilon}\right)^{\nu-1}\left(
\frac{r}{\epsilon^2\sqrt{r^2+\epsilon^2}} \right)\,.
\end{equation}
Notice that for $p=2$,
\begin{equation}\label{A0p2}
A^0(r) = \frac{e}{\epsilon}
\left(\frac{1}{\sqrt{1+\frac{r^2}{\epsilon^2}}} +
\frac{1}{2\left(1+\frac{r^2}{\epsilon^2}\right)^{\frac{3}{2}}}
\right) ,\quad e=\frac{e^{(7)}}{R^2\epsilon}\,,
\end{equation}
which becomes the Coulomb potential for $r\gg\epsilon$ and its
finite at the 3D source position:
\begin{equation}
\lim_{r\rightarrow 0}A^0(r)=\frac{3e}{2\epsilon}.
\end{equation}
The source self-energy is now
\begin{equation}
E_{self}^{(p=2)}=\frac{315e^2}{16384\pi\epsilon}.
\end{equation}

\begin{figure}[htb]
\center{\includegraphics[height=7cm,width=9cm]{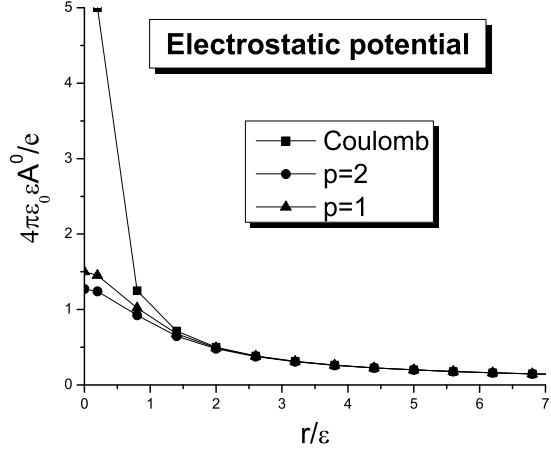}}
\caption{Electrostatic potential of the point particle for the standard 4D Coulomb
case and $p=1,2$.}
\label{coulom}
\end{figure}

The static potentials for $p=1$, $p=2$ and Coulomb's are compared in
Fig. (\ref{coulom}). Remarkably as we have mentioned, the
electrostatic potentials corrected by the extra dimensions are
finite at the 3D position of the charge.

It is interesting and natural to explore possible consequences of
the modified electrostatic potentials we just obtained using known
experiments like the Cavendish and scattering ones. We do so in the
following section.

\section{Phenomenology of the electrostatic potential}

\subsection{Cavendish experiment}\label{SubsecCavendish}
From the different results obtained to verify the accuracy of the
electrostatic force, we have chosen the ones obtained by Plimpton
and Lawton \cite{Plimpton:1936} and more recent modifications (see
\cite{Tu:2005ge,0026-1394-41-5-S04} for a review of the different
experiments). The reason is that this belongs to a series of
experiments in which the main idea was to test the accuracy of
Coulomb's force between charged particles using similar techniques
as the one used by Cavendish to test the gravitational force (see
for instance \cite{Goldhaber:2008xy} for a recent review on the
different perspectives and experiments performed to test different
aspects of electrodynamics). In the modern version of Cavendish
experiment we have a modified electromagnetic potential for a charge
$Q$
\[
V= V_C + \delta V,
\]
where $\delta V$ is the modification    to  the   Coulomb potential.
The idea behind the concentric charged spheres  experiments  is that
only for the Coulomb potential the interior of a charged sphere
is field  free and therefore the potential there is a constant. Then
the potential difference between a charged outer sphere and the
uncharged inner sphere is zero only if the potential is that of
Coulomb.  Any deviation from this would imply a nonvanishing
potential difference between the spheres that can be measured.

The potential of a sphere with a charge $Q$ and radius $c$ at a
distance $r$ from the center is
\[
U(Q,r,c)= \frac{Q}{2cr}\left[ f(r+c) - f(|r-c|)\right] ,
\]
where the function $f$ is given by
\[
f(r) = \int_0^r {ds \;s V(s,Q=1)},
\]
It is easy to verify that for $V=\alpha/r$, $U(Q,r<c,c) =const.$,
that is the potential is constant in the interior.

In the simplest version of the Cavendish experiment one has an outer
sphere of radius $b$, charged to a certain voltage, and then
measures the relative voltage difference to the uncharged inner
sphere of radius $a<b$,
\begin{equation}
\label{gamma1} \gamma_{ab}=\left|\frac{{\mathcal V}_{b}-{\mathcal
V}_{a}}{{\mathcal V}_{b}}\right|
\;\;=\;\;\left|\frac{U(Q,b,b)-U(Q,a,b)}{U(Q,b,b)}\right|.
\end{equation}
Plimpton and Lawton found that $|\gamma_{ab} | \le  3 \times
10^{-10}$ with $a= 0.696$ m, $b=0.762$m. Here we will calculate
$|\gamma_{ab} |$  for the two potentials corresponding to $p=1$ and
$p=2$, namely
\begin{eqnarray}
v_1&=& \frac{2Q}{\epsilon \pi}\left( \frac{1}{1+
\frac{r^2}{\epsilon^2}}+
\frac{\arctan(\frac{r}{\epsilon})}{\frac{r}{\epsilon}}  \right),\nonumber \\
v_2&=&\frac{Q}{\epsilon }\left( \frac{1}{      \sqrt{   1+
\frac{r^2}{\epsilon^2}  }    } +   \frac{1}{2 (1 +
\frac{r^2}{\epsilon^2} )^{\frac{3}{2}}}  \right).
\end{eqnarray}
The results, to first order in $\epsilon$ are
\begin{equation}
|\gamma_{1ab} |= \frac{\epsilon}{ \pi b},   \;\;  |\gamma_{2ab} |=
\frac{\epsilon}{ 4 b}.
\end{equation}
Taking into account the experimental bound of Plimpton and Lawton
this means that $\epsilon \le 7.18 \times 10^{-10} m $ or $\epsilon
\le 9.14 \times 10^{-10} m $   for $p=1$ and $p=2$, respectively.

The more recent version of the Cavendish experiment employs four
concentric spheres of radii a, b, c, d  in increasing order. The
Outer sphere has a charge $Q$ and the next one -$Q$. Then the potential
at radius r is given by
\begin{equation}
U(Q,r,c,d)= \frac{Q}{2dr}\left[ f(r+d) - f(|r-d|)\right]-
\frac{Q}{2cr}\left[ f(r+c) - f(|r-c|)\right],
\end{equation}
The experiment sets a bound for the ratio of the potential
differences between the two uncharged spheres and the two outer
spheres
\begin{equation}
\gamma_{abcd}=\left|\frac{{\mathcal V}_{b}-{\mathcal
V}_{a}}{{\mathcal V}_{c}-{\mathcal V}_{d}}\right|= \left|
\frac{U(Q,b,c,d)-U(Q,a,c,d)}{U(Q,c,c,d)-Q(Q,d,c,d)}\right|.
\end{equation}
Williams et al.  \cite{Williams:1971ms} found  that $|\gamma_{abcd}
| \le  2 \times 10^{-16}$ with $a= 0.60$ m, $b=0.94$m, $c= 0.947$m
and $d= 1.27$m. Here we will calculate $|\gamma_{abcd} |$ for the
two potentials corresponding to $p=1$ and $p=2$, and using the
experimental limits to constrain $\epsilon$. A straightforward
calculation gives, to leading order in $\epsilon$
\begin{equation}
\gamma_{1abcd}= \frac{c d
   \left(\frac{4 (c-d) (c+d)
   \left(-2
   a^2+c^2+d^2\right)}{\left(a
   ^2-c^2\right)^2
   \left(a^2-d^2\right)^2}+
   \frac{4 \left(2 b^2 (c-d)
   (c+d)-c^4+d^4\right)}{\left
   (b^2-c^2\right)^2
   \left(b^2-d^2\right)^2}
   \right)}{3 \pi
   (c-d)}  \epsilon ^3 +O\left(\epsilon
   ^4 \right),
\end{equation}
\begin{equation}
\gamma_{2abcd}= \frac{c d
   \left(\frac{\frac{\frac{1}{
   (a+c)^3}+\frac{1}{(a-c)^3}}
   {c}+\frac{\frac{1}{(d-a)^3}
   -\frac{1}{(a+d)^3}}{d}}{a}+
   \frac{\frac{\frac{1}{(c-b)^
   3}-\frac{1}{(b+c)^3}}{c}+
   \frac{\frac{1}{(b+d)^3}+\frac
   {1}{(b-d)^3}}{d}}{b}\right)
   }{16 (c-d)} \epsilon ^4   +O\left(\epsilon
   ^5\right).
\end{equation}

Taking into account the experimental value obtained by    Williams
et al. ($|\gamma_{abcd} | \le  2 \times 10^{-16}$ ) the
corresponding bounds for $\epsilon$ are $\epsilon \le 2.80  \times
10^{-7} m $ or $\epsilon \le 4.02 \times 10^{-6} m $  for $p=1$ and
$p=2$, respectively. In this case the two sphere experiment gives a
tighter constraint on  $\epsilon$. The reason for this may be the
peculiarities of the modification of the Coulomb potentials that in
our case contains positive powers of $r$.

\subsection{Scattering by Helium atoms} \label{SubsecHelium}

We shall study the collision of a particle of charge $ze$ and mass
$m$ with an atom of atomic number $Z$. Notice that an exact
formulation of this problem requires the use of a many-body
Hamiltonian which describes all the particles of the system, however
we shall make the assumption that the complicated interaction of the
incident particle with the constituents of the atom can be
accounted for by an effective electrostatic potential $V(r)$ in which
the incident particle travels.

\begin{figure}[htb]
\center{\includegraphics[height=7cm, width=9cm]{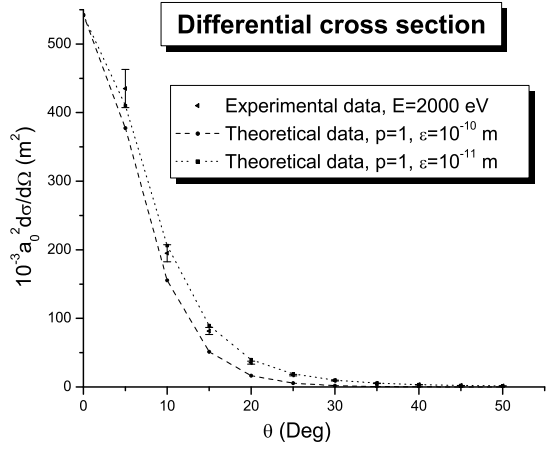}} \caption{
Comparison of experimental differential cross section
\cite{0022-3700-9-2-009} with that corresponding to one compact dimension, Eq.
(\ref{eq:csp1}).} \label{sep1}
\end{figure}

\begin{figure}[htb]
\center{\includegraphics[height=7cm, width=9cm]{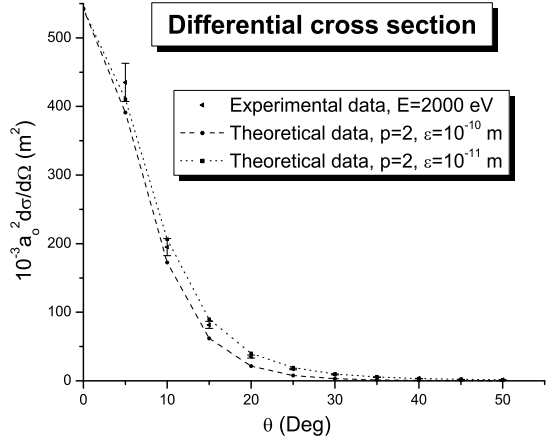}} \caption{
Comparison of experimental differential cross section
\cite{0022-3700-9-2-009} with that corresponding to two compact dimensions, Eq.
(\ref{eq:csp2}).} \label{sep2}
\end{figure}

It is physically reasonable that the electrostatic potential in
which the incident particle travels is well approximated by
\begin{equation}\label{ec.potential}
V(\vec{r})=ze\left[Zev_{1,2}(\vec{r})+e\int\rho(\vec{r}')v_{1,2}\left(|\vec{r}-\vec{r}'|\right)d^3\vec{r}'\right],
\end{equation}
were $\vec{r}$ is the position vector of the incident particle and
$v_{1,2}(\vec{r})$ are given by (\ref{A0p1}) and (\ref{A0p2}). The
first term is due to the field of the nucleus and the second term is
the potential of the atomic electrons, described in terms of an
effective electron density $\rho$. It is worth mentioning that in
this description we are neglecting all effects of symmetry and spin.
For neutral atoms, the density satisfies
\begin{equation}
\int\rho(\vec{r})d^3\vec{r}=Z.
\end{equation}
When the incident particle carries sufficiently high energy, the
scattering amplitudes can be easily evaluated by the Born
approximation
\begin{equation}\label{Born}
f(\theta)=-\frac{m}{2\pi\hbar^2}\int e^{i\vec{q}\cdot
\vec{r}}V(\vec{r})d^3\vec{r},
\end{equation}
where $\vec{q}=\vec{k}_0-\vec{k}$, and $\vec{k}_0$ and  $\vec{k}$ are the initial and final momentum, respectively. Since the scattering
is elastic, $|\vec{k}|=|\vec{k}_0|=k$. Thus introducing Eq.
(\ref{ec.potential}) in (\ref{Born}) and making the following change
of variable $\vec{R}=\vec{r}-\vec{r}'$ we have
\begin{eqnarray}
f_{1,2}(\theta)&=&-\frac{me^2}{2\pi\hbar^2}\left[zZ\int
e^{i\vec{q}\cdot \vec{r}} v_{1,2}(\vec{r})d^3\vec{r} -z\int
e^{i\vec{q}\cdot
\vec{R}}v_{1,2}\left(\vec{R}\right)d^3\vec{R}\int\rho(\vec{r}')e^{i\vec{q}\cdot
\vec{r}'}d^3\vec{r}'\right]
,\nonumber\\
&=&-\frac{me^2z}{2\pi\hbar^2}\left[Z -F(\vec{q})\right]\int
e^{i\vec{q}\cdot \vec{r}} v_{1,2}(\vec{r})d^3\vec{r} ,
\end{eqnarray}
$F(\vec{q})$ is called the form factor of the atom. We defined
$F(\vec{q})$ as
\begin{equation}\label{ec.amplitude}
F(\vec{q})=\int\rho(\vec{r})e^{i\vec{q}\cdot \vec{r}}d^3\vec{r}.
\end{equation}
When the potential is spherically symmetric, the angular integration
can be performed to give
\begin{equation}
f_{1,2}(\theta)= -\frac{2me^2z}{\hbar^2}\left[Z
-F(\vec{q})\right]\int_0^{\infty}
\frac{\sin(qr)}{qr}v_{1,2}(r)r^2dr,
\end{equation}
with $q=|\vec{q}|=2k\sin(\theta/2)$ and $r=|\vec{r}|$. The
evaluation of this integral depends on the form that $v_{1,2}(r)$
takes. We first calculate $f_1(\theta)$
\begin{equation}
f_{1}(\theta)= -\frac{4me^2z}{\pi\hbar^2}\frac{1}{\epsilon q}\left[Z
-F(\vec{q})\right]\int_0^{\infty}\sin(qr) \left(
\frac{1}{1+\frac{r^2}{\epsilon^2}}+\frac{\arctan\left(\frac{r}{\epsilon}\right)}{\frac{r}{\epsilon}}
\right) rdr,
\end{equation}
using the following relations (see \cite{0122947606})
\begin{eqnarray}
\int_0^{\infty}\frac{\sin(qr)}{1+\frac{r^2}{\epsilon^2}}rdr&=&\frac{\pi}{2}\epsilon^2e^{-q\epsilon},\\
\int_0^{\infty}\sin(qr)
\frac{\arctan\left(\frac{r}{\epsilon}\right)}{r}rdr&=&\frac{\pi}{2}\frac{e^{-q\epsilon}}{q},
\end{eqnarray}
$f_1(\theta)$ can be written as
\begin{equation}
f_{1}(\theta)=-
\frac{2me^2z}{\hbar^2}\left[Z-F(\vec{q})\right]\left[\frac{1}{q^2}+\frac{\epsilon}{q}\right]e^{-q\epsilon}.
\end{equation}
Considering the form of $v_2(r)$, $f_2(\theta)$ can be expressed as
\begin{equation}
f_{2}(\theta)= -\frac{2me^2z}{\hbar^2}\frac{1}{\epsilon q}\left[Z
-F(\vec{q})\right]\int_0^{\infty}\sin(qr) \left(
\frac{1}{\sqrt{1+\frac{r^2}{\epsilon^2}}}+\frac{1}{2\left(1+\frac{r^2}{\epsilon^2}\right)^{\frac{3}{2}}}
\right) rdr.
\end{equation}
Now let us consider the integrals
\begin{eqnarray}
\int_0^{\infty}\frac{\sin(qr)}{\sqrt{1+\frac{r^2}{\epsilon^2}}}rdr&=&\epsilon^2K_1(q\epsilon)
,\\
\int_0^{\infty}\frac{\sin(qr)}{\left(1+\frac{r^2}{\epsilon^2}\right)^{\frac{3}{2}}}rdr&=&\epsilon^3qK_0\left(q\epsilon\right),
\end{eqnarray}
where $K_0(x)$ and $K_1(x)$ are Bessel functions of zeroth and first
order, respectively. Thus $f_2(\theta)$ takes the form
\begin{equation}
f_{2}(\theta)= -\frac{2me^2z}{\hbar^2}\left[Z -F(\vec{q})\right]
\left[ \frac{\epsilon}{q}K_1(q\epsilon)+\frac{\epsilon^2}{2}
K_0(q\epsilon) \right].
\end{equation}
For Helium we can calculate the electron density as
\begin{equation}
\rho(r)=Z\left(\frac{b^3}{\pi a_0^3}\right)e^{\frac{-2br}{a_0}},
\end{equation}
with $b$ being the effective charge and having the value 1.69 for Helium while $a_0$ is
the Bohr radius. The form factor becomes
\begin{equation}
F(q)=\frac{Z}{\left(1+\frac{a_0^2q^2}{4b^2}\right)^2}.
\end{equation}
The differential scattering cross section for elastic processes thus
become
\begin{eqnarray}
\left(\frac{d\sigma}{d\Omega}\right)^{(p=1)}&=&
\left(\frac{2zZ}{a_0q^2}\right)^2\left[1-\frac{1}{\left(1+\frac{a_0^2q^2}{4b^2}\right)^2}\right]^2\left[1+q\epsilon\right]^2e^{-2q\epsilon},\label{eq:csp1}\\
\left(\frac{d\sigma}{d\Omega}\right)^{(p=2)}&=&
\left(\frac{2zZ}{a_0q^2}\right)^2\left[1-\frac{1}{\left(1+\frac{a_0^2q^2}{4b^2}\right)^2}\right]^2
\left[ q\epsilon K_1(q\epsilon)+\frac{q^2\epsilon^2}{2}
K_0(q\epsilon) \right]^2\label{eq:csp2}.
\end{eqnarray}
For incident electrons, we set $z=-1$ and $Z=2$ for the Helium atom.
To complete the analysis we compare the theoretical results with
the corresponding experimental ones. This comparison is made explicit in Figures (\ref{sep1}) and
(\ref{sep2}). For both $p=1,2$ a best agreement is attained when $\epsilon \sim 10^{-10}m$.


\section{Discussion}\label{conclusions}

The ever increasing accuracy with which electrodynamics has been
tested naturally lends itself to consider it as a probe to set
bounds for possible deviations coming from the existence of extra
dimensions. Amongst different models the so called Randall-Sundrum
ones including a single 3-brane and $p$ extra compact dimensions
(RSII-$p$) have provided simple scenarios that yield effects well
under control. Take for example the Casimir force
\cite{Linares:2010uy,Frank:2008dt}: In a nutshell the field modes
corresponding to the extra dimensions add up to modify the usual
Casimir force expression and the deviations are assumed to be bounded by the uncertainties in the experimental data. This in turn sets bounds for the parameters of the brane model.

In this work we have explored the static potential produced by a
scalar and a charged sources, respectively, in RSII-$p$. These
sources are {\em effectively pointlike} from the perspective of an
observer sitting in the usual 3D space. However they stretch
uniformly along the $p$ compact dimensions thus having the structure
of a $T^{p}$ torus. Remarkably the effective potentials turn out to
be non-singular at the position in 3D space. At first one may think
this is related to the fact the sources stretch along the extra
dimensions, similarly as in models of charged spherical shells
\cite{rohrlich}. This is not the case as a more careful look
reveals: the potential produced by either a charged
ring or a torus is not finite at the source itself
\cite{Andrews,Kondratev,Bannikova:2010qw}. The RSII-p scenario thus
allows to regularize the 3D potentials and selfenergies. Indeed the
combined limit having AdS radius and compact size going to zero yields the usual standard divergent result.

We have determined the potentials in the low energy regime in terms
of light modes; this entails approximating the continuous modes
given in terms of Bessel functions by their small argument form
whereas for the compact modes we keep the zero mode only. Within
this approximation a delicate balance occurs between part of the
massive sector contribution to the potential and the zero mode.
Since the zero mode is responsible for the usual singular $1/r$ term, the potential characteristic of massless fields, the balance just
described regularizes such a divergence. Moreover the remaining
effective potential becomes the usual $1/r$ within a few times
$\epsilon$ away from $r=0$ and provides finite selfenergies as determined from the usual 3D formulae.

To probe the effective potentials we proposed to consider two types
of experiments. First we adopted
the long known Cavendish experiment with two and four conducting
spheres that is used to test the form of the Coulomb force. To be
consistent with know experimental results for the case of two
spheres a value of $\epsilon \sim 10^{-10}m$ is required. The four
spheres setting however turns out to produce a milder bound
$\epsilon \sim 10^{-7}m$, probably due to the positive powers of the
correcting terms of the effective potentials when developing around
$1/r$. The second possibility we studied to test our effective
potentials was to consider electrons scattered off by Helium. A
comparison of the differential cross section modified by the RSII-p
scenario with the curve fitting experimental data indicates
consistency with a value of $\epsilon \sim 10^{-11}m$. In a previous
work \cite{MoralesTecotl:2006eh} we used the Lamb shift to set a
bound of $\epsilon \sim 10^{-14}m$ for $p=1$, and $\epsilon \sim
10^{-13}m$, for $p=2$, which clearly are stronger than the ones obtained in the present work.

The fact that for both the scalar and electromagnetic case the
potentials become well behaved leads naturally to the question of
whether the same results holds for the gravitational case. This is
work under study and will be reported elsewhere.
Indeed, historically, finiteness of the potentials have led in the past to the idea that gravity regulates the self-energy of the charged point particle \cite{Arnowitt:1962hi} as well as nonlinear field equations to achieve the finiteness of the electric field \cite{Born:1934gh}.

In the low energy
regime we have focused on in  this work there are some other
possible directions which can be pursued. These include a reanalysis
of the radiation reaction problem in both electromagnetic an
gravitational cases as well as the complete understanding of the
regularization of the potentials and in particular its relation to
the topology of the sources together with their dimensionality.

\begin{acknowledgments}
HAMT acknowledges partial support from grant CONACyT-NSF Strong backreaction effects in quantum cosmology.
\end{acknowledgments}

\bibliography{bibliography}

\end{document}